# Material Made of Artificial Molecules and Its Refraction Behavior under Microwave

Tao Zhang

*College of Nuclear Science and Technology, Beijing Normal University, Beijing 100875, China*

(taozhang@bnu.edu.cn)

**Abstract** Focal-plane imaging with microwave and millimeter wave has many potential applications. The bad resolution is a main shortcoming of this technique at present. In this paper, influence of frequency of electromagnetic wave on the refractive energy is analyzed. According to the refractive mechanism of electromagnetic induction, the principle for designing artificial materials with enough refractive ability is proposed. The material of artificial molecules is fabricated with cage-shaped granules of conductor (CGC). The refractive index of CGC material to microwave was measured by assembling CGC prism with PS (polystyrene) prism or SGC (solid granule of conductor) prism. The assemblies greatly eliminate displacement of the outgoing microwave caused by diffraction. This is a key point to achieve direct measurement of the refractive index under the microwave. CGC material shows a considerable refractive ability. SGC material and PS show little refractive ability. The results are useful in promoting image technique of microwave and millimeter-wave by improving the refractive ability and eliminating detriment of the diffraction on the lenses.

**Key word**: material; artificial molecule; microwave; refraction; diffraction; measurement; imaging

## 0  Introduction

Focal-plane imaging technique by microwave and millimeter wave is an application based on electromagnetism and optics, and has been investigated extensively in recent years [1, 2]. Its potential applications include landing system, security check, medical imaging, and so on. The lenses play a pivotal role in this technique because their refractive performance greatly influences the resolution of the images. So far, bad resolution of the lenses limits the applications of this imaging technique. Breakthroughs in the theory and the technology are expected to solve these problems. In this paper, theoretical analyses on the refraction of microwave are presented, verifying experiments are conducted. The results are irradiative.

## 1  Theory

The refractive mechanism of electromagnetic induction and the concept of the refractive energy have been put forward in previous papers [3, 4]. The mechanism suggests that the refraction phenomena result from the electromagnetic induction between electromagnetic waves (EWs) and electrons or other charged particles. The formulae of refractive index have been derived with the mechanism. The mechanism has been used to study the interactions between the light and the electrons, and have made the following results: The influences of static electric field and static magnetic field on the refractive index, the anisotropy of the refractive index, and the reason for $n_{H2}>n_{HE}$ and $n_{N2}>n_{O2}$ ($n_{H2}$, $n_{HE}$, $n_{N2}$ and $n_{O2}$ are the refractive indices of $H_2$, He, $N_2$ and $O_2$, respectively) have been explained with the mechanism. All these results support the mechanism. Reference [4] indicates that the refractive energy has a quantum characteristic. This means that energy of EW may become the refractive energy of the electron only when the interaction of electromagnetic induction between the EW and the electron is strong enough. In other words, material will not have refractive ability, or its refractive index =1, to the EW which frequency is below certain critical value. With this point of view, we reason out the behavior of refractive ability of molecules vs frequency of EW: (1) In the range of light frequency, the electromagnetic induction between light and the electrons in molecules is strong enough, and all molecules have enough refractive ability; (2) As the frequency decreasing, firstly the inner electrons and then the outer electrons in molecules gradually do not produce refractive energy, and molecules lose their refractive ability gradually; (3) In the range of microwave frequency, the electrons in most molecules do not produce refractive energy, and molecules and dielectrics made of them in nature have little refractive ability. It should be noted that diffraction phenomena exist in all the three bands of frequency.

The artificial materials had been developed in order to decrease the weights and increase the refractive index of the lenses used under microwave. But the refractive indices of these conventional artificial materials are not high enough to fulfil the expectation [5]. We believe that the conductor pieces in the conventional artificial materials only offer a limited refractive ability because the conductor pieces are solid. Compared with the solid conductor piece, a cage-shaped granule of conductor (CGC for short) is supposed to show a higher refractive ability. The artificial materials made of CGCs should have enough refractive indices. The foundation for above idea is as follow: There are closed loops of conductor in CGC. The magnetic field of EW can go through these closed loops, so that strong interactions of electromagnetic induction can be produced in CGC, see Fig.1. This characteristic makes CGC have enough refractive ability, while solid conductor piece can not produce strong interaction of electromagnetic induction. In a word, as for designing artificial material of high refractive index, it should be kept in mind that a conductor piece in the artificial material may contribute enough refractive ability only when it is able to encircle adequately the magnetic field lines of EWs.

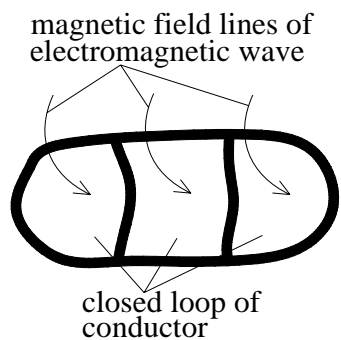

Fig. 1 Segment of CGC. Three closed loops of conductor on CGC are shown. The magnetic field lines of electromagnetic wave go through the closed loops.

## 2  Experimental methods

The structure of single CGC is shown in Fig. 2. Like the interaction of electromagnetic induction between molecules and light, a strong interaction of electromagnetic induction can be produced between CGC and microwave. So CGC is a kind of artificial molecule. Large numbers of CGCs are distributed uniformly in space and are isolated from each other by dielectric material, thus form the artificial molecular material — the CGC material. The conventional artificial material is made with solid granules of conductor (SGC) for comparison. Prisms are made with the CGC material, the SGC material and polystyrene (PS), respectively. The shape of the prisms is shown in Fig. 3. The end of the prism is in a shape of right triangle with one acute angle being 21°. The volume fraction of the granules in the CGC material is the same as that in the SGC material.

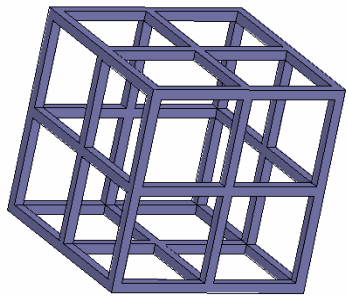

Fig. 2 Sketch map of CGC. CGCs are electrically isolated from each other in the CGC material.

The microwave with a frequency of 2.3 GHz is emitted from a horn. The prisms are put at the open of the horn, see Fig. 4. There are 5 kinds of manners for putting the prisms: (1) CGC prism alone (Fig. 4(a)), (2) PS prism alone (Fig. 4(b)), (3) SGC prism alone (Fig. 4(b)), (4) CGC prism and PS prism (Fig. 4(c)), and (5) CGC prism and SGC prism (Fig. 4(c)). In all the 5 kinds of manners the 21° angle of CGC prism points towards the positive-angle side, and the 21° angles of PS prism and SGC prism point towards the negative-angle side. In manners (4) and (5) the two prisms make assemblies in a shape of box with its front face and back face being parallel to each other. Move the detector along the circle (see Fig. 4) to measure the microwave signal. The horn locates at the centre of the circle.

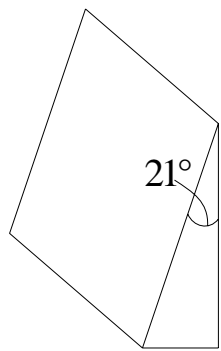

Fig. 3 Sketch map of the prisms.

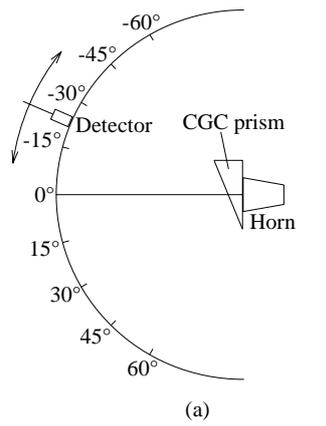

(a)

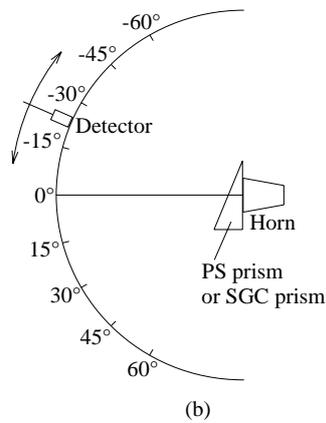

(b)

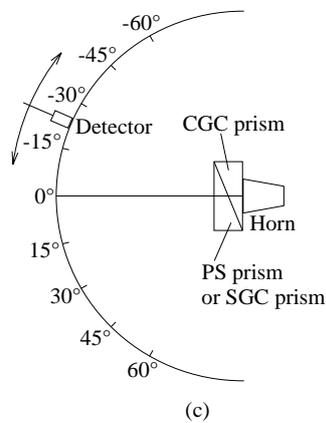

(c)

Fig. 4 Experimental setup.

## 3  Experimental results and discussions

The experimental results are shown in Fig. 5. The curve is roughly symmetrical with its peak locates at 0° when no prism is put at the horn. The curve peak moves towards the negative-angle side when putting CGC prism alone at the horn, and the curve peaks move towards the positive-angle side when putting PS prism alone or CGC prism alone at the horn. All the three kinds of prisms make the outgoing microwave deflect. When put the assemblies of the prisms at the horn, the curve peaks move towards the negative-angle side with a displacement angle of 11° for CGC/PS assembly and a displacement angle of 9° for CGC/SGC assembly. These results indicate that CGC prism has a stronger refractive ability than PS prism or SGC prism.

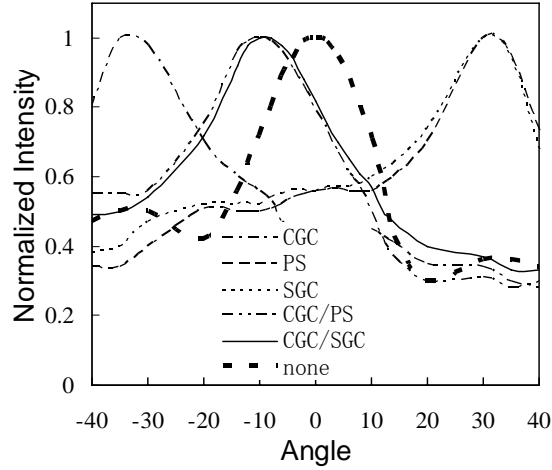

Fig. 5 The experimental results. CGC: CGC prism alone; PS: PS prism alone; SGC: SGC prism alone; CGC/PS: assembly of CGC and PS prisms; CGC/SGC: assembly of CGC and SGC prisms; none: no prism.

At present it is believed that the refractive index $n_{PS}$ of PS is about 1.5 under microwave. We calculate the refractive index $n_{CGC}$ of the CGC material in the assembly of CGC/PS prisms (see Fig. 6) with $n_{PS}=1.5$. The result is $n_{CGC}=2.01$. This value is too large to be reasonable. So $n_{PS}$ probably is not as large as expected. We believe that $n_{PS}$ is close to 1 under the microwave according to the analysis in above section. The displacement of the curve peak by PS prism alone results only from diffraction on PS prism, not from refraction. In other words, PS prism only has diffractive ability, and almost no refractive ability, while CGC prism has both refractive ability and diffractive ability. In the assembly of CGC/PS prisms, as the front face and the back face of the assembly are parallel to each other, and the microwave goes through the interfaces in an appropriate angle, diffraction should not cause evident displacement of the curve peak. Therefore in the assembly the displacement is caused mainly by refraction on CGC prism. Following this judgement, omitting refractive ability of PS prism, $n_{CGC}$ is calculated with the deflection angle of 11°. The result is $n_{CGC=}1.48$. The calculated refractive index of this CGC material is 1.46 (The calculating method will be presented in another article).

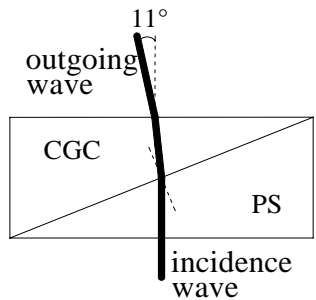

Fig. 6 Propagation of the microwave in the assembly of CGC prism and PS prism.

Direct measurement according to refractive principle is supposed to produce reliable data of refractive indices. Direct measurement of the refraction of light can be conducted with little difficulty. In the range of microwave and millimeter wave, however, direct measurement of the refractive indices is difficult because of the trouble of diffraction. By eliminating the displacement effect caused by diffraction, this paper offers a feasible method for the direct measurement.

In the cases of both SGC prism alone and the assembly of CGC/SGC prisms, behavior of SGC prism is similar to that of PS prism, see Fig. 5. This result indicates that the refractive ability of SGC prism is very weak, and the displacement of the curve peak by SGC prism alone mainly results from diffraction also. Both CGC prism and SGC prism are made of artificial materials. The conductor pieces in SGC prism are solid and the conductor pieces in CGC prism are in close-loop structure. Therefore CGC can produce strong electromagnetic induction under microwave while SGC can not, see Fig. 7. This difference brings on their different refractive abilities.

The displacement angles of the outgoing microwave for CGC prism alone is 35°, for PS prism alone is 31°, and for SGC prism alone is 30°, see Fig. 5. These displacement angles should mainly result from diffraction. From above text we know that refraction results in 9°- 11° displacement angles in our experiments. So the displacement effect of diffraction is much stronger than that of refraction sometimes. CGC prism has an extra deflective effect from refraction. But its displacement angle (35°) does not show big margin over those (31° and 30° respectively) of PS prism and SGC prism. This result indicates that the displacement effect of diffraction has an overwhelming dominance over that of refraction in this work.

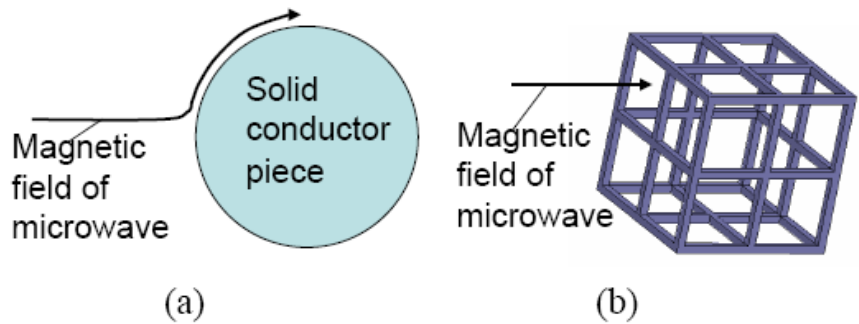

Fig. 7 (a) Magnetic field of microwaves can not go through the solid conductor piece in SGC material — a kind of conventional artificial materials. (b) Magnetic field of microwaves goes through the conductor piece in CGC material and produces strong electromagnetic induction.

Although diffraction sometimes has stronger displacement effect than refraction, and diffraction may focus, one prefers focusing by refraction to diffraction in the imaging technique of microwave and millimeter wave. Here the reason is that the quality of focus by diffraction is not as good as that by refraction. Generally, diffraction phenomena are inevitable when microwave and millimeter wave shoot on the lenses. If there are simultaneously diffraction focuses and refraction focuses in the image side, quality of the image will be bad. To eliminate this harm one need to avoid the diffractive displacements of the outgoing microwave and millimeter wave. Currently the lenses used for focusing microwave and millimeter wave are commonly made of polymers, and have considerable thicknesses. According to above analysis, it is clear that diffraction contributes a lot to the focusing ability of these lenses, and does detriment to the image quality. Measures for correcting this flaw may be taken in two ways: (1) Fabricate the lenses with the CGC material to increase their refractive indices, so as to enhance their refraction focusing; (2) Adopt appropriate outer contours of the lenses to avoid the diffractive displacement, so as to weaken the diffractive focusing, see Fig. 8. Some materials, such as PS, have little refractive ability under microwave (and millimeter wave). This characteristic facilitates the measure (2): One may use the lacking-refractive-ability materials to change the outer contours of the lenses, eliminating the diffractive displacement in the same time without influencing the refractive behaviors of the lenses.

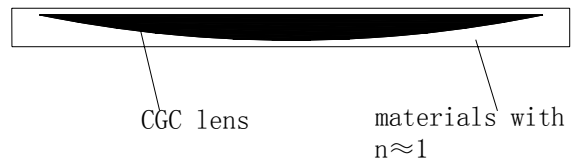

Fig. 8 Assembled lens with an outline in slab shape.

## 4 Conclusion

Direct measurement of the refractive index to microwave is realized with the method of combining two prisms into an assembly in a box shape. The CGC material shows a considerable refractive ability. On the contrary, the SGC material and PS show little refractive ability. The results confirm our theory for designing artificial materials of high refractive indices. Attention should be paid to eliminate the displacement effect by the diffraction when measuring the refractive indices under microwave. The results in this paper indicate that when adopting an appropriate outer contour by appending lacking-refractive-ability materials to an object, the diffractive displacement can be weakened without altering the refractive behavior of the object. This idea is useful in avoiding the harms of diffraction in both the measurement of refractive indices and the imaging technique with microwave and millimeter wave.

## Acknowledgments

Supported by Beijing Science Technology New Star Program (Grant No. 952870400).